\documentclass[%
 aip,
 jmp,%
 amsmath,amssymb,
 reprint,%
]{revtex4-1}

\usepackage{graphicx}% Include figure files
\usepackage{dcolumn}% Align table columns on decimal point
\usepackage{bm}% bold math
\usepackage{color}% bold math
\begin{document}
\title{Preserving the 7$\times$7 surface reconstruction of clean Si(111) by graphene adsorption}

\author{Justin C. Koepke}
\affiliation{Department of Electrical and Computer Engineering, Beckman Institute for Advanced Science and Technology, and Micro and Nanotechnology Laboratory. University of Illinois at Urbana-Champaign. Urbana, IL 61801. USA}
\author{Joshua D. Wood}
\affiliation{Department of Electrical and Computer Engineering, Beckman Institute for Advanced Science and Technology, and Micro and Nanotechnology Laboratory. University of Illinois at Urbana-Champaign. Urbana, IL 61801. USA}
\author{Cedric M. Horvath}
\affiliation{Department of Physics. University of Arkansas. Fayetteville, AR 72701. USA}
\author{Joseph W. Lyding}
\affiliation{Department of Electrical and Computer Engineering, Beckman Institute for Advanced Science and Technology, and Micro and Nanotechnology Laboratory. University of Illinois at Urbana-Champaign. Urbana, IL 61801. USA}
\author{Salvador Barraza-Lopez}
\affiliation{Department of Physics. University of Arkansas. Fayetteville, AR 72701. USA}
\email{sbarraza@uark.edu}

\date{\today}% It is always \today, today,

\begin{abstract}
We employ room-temperature ultrahigh vacuum scanning tunneling microscopy (UHV STM) and {\em ab-initio} calculations to study graphene flakes that were adsorbed onto the Si(111)$-$7$\times$7 surface. The characteristic 7$\times$7 reconstruction of this semiconductor substrate can be resolved through graphene at all scanning biases, thus indicating that the atomistic configuration of the semiconducting substrate is not altered upon graphene adsorption. Large-scale {\em ab-initio} calculations confirm these experimental observations and point to a lack of chemical bonding among interfacial graphene and silicon atoms. Our work provides insight into atomic-scale chemistry between graphene and highly-reactive surfaces, directing future passivation and chemical interaction work in graphene-based heterostructures.
\end{abstract}

\maketitle

Graphene is a single layer of sp$^2$-bonded carbon atoms\cite{RMP,KatsnelsonBook} that has an unusually large carrier mobility and thermal conductivity.~\cite{r1,r2,r3} Its electronic structure is represented by a conical, relativistic-like energy-momentum relation at energies within 1 eV from its Fermi level. Silicon, on the other hand, remains the workhorse of present electronic technologies, forming a stable (111)$-$$7\times 7$ reconstructed surface after high temperature treatment.\cite{Binnig,Himpsel,Takayanagi,Brommer,Pandey,Himpsel2,Stroscio,Ancilotto} The Si(111)$-$$2\times 1$ surface reconstruction\cite{Pandey,Himpsel2,Stroscio,Ancilotto} can only be produced by cleavage,\cite{Spence} giving way to the $7\times 7$ reconstruction at 900 K.

Many studies have addressed graphene's electronic, mechanical, and topographic properties upon adsorption onto semiconducting\cite{E5,E6,E8,NanoLetters} and insulating\cite{r1,SC1,SC2,SC3,Korean,hbnPRL,hbnLeRoy,hbn1,hbn2,hbn3,hbn4} substrates. Still, clean Si(111) is perhaps the most reactive semiconductor to be interfaced with graphene or another inert two-dimensional material to date. It has a high density of dangling bonds, larger topographic modulation than Si(100)$-$2$\times$1 or GaAs(110), and higher crystalline ordering over SiO$_2$ or graphitized SiC. While the properties of graphene adsorbed on Si(111) were computationally examined,\cite{PRL2013} we aim to provide a comprehensive experimental and computational characterization of the intriguing graphene-Si(111) system, thereby elucidating interfacial chemistry among graphene and reactive surfaces.

\begin{figure*}[tb]
\includegraphics[width=0.8\textwidth]{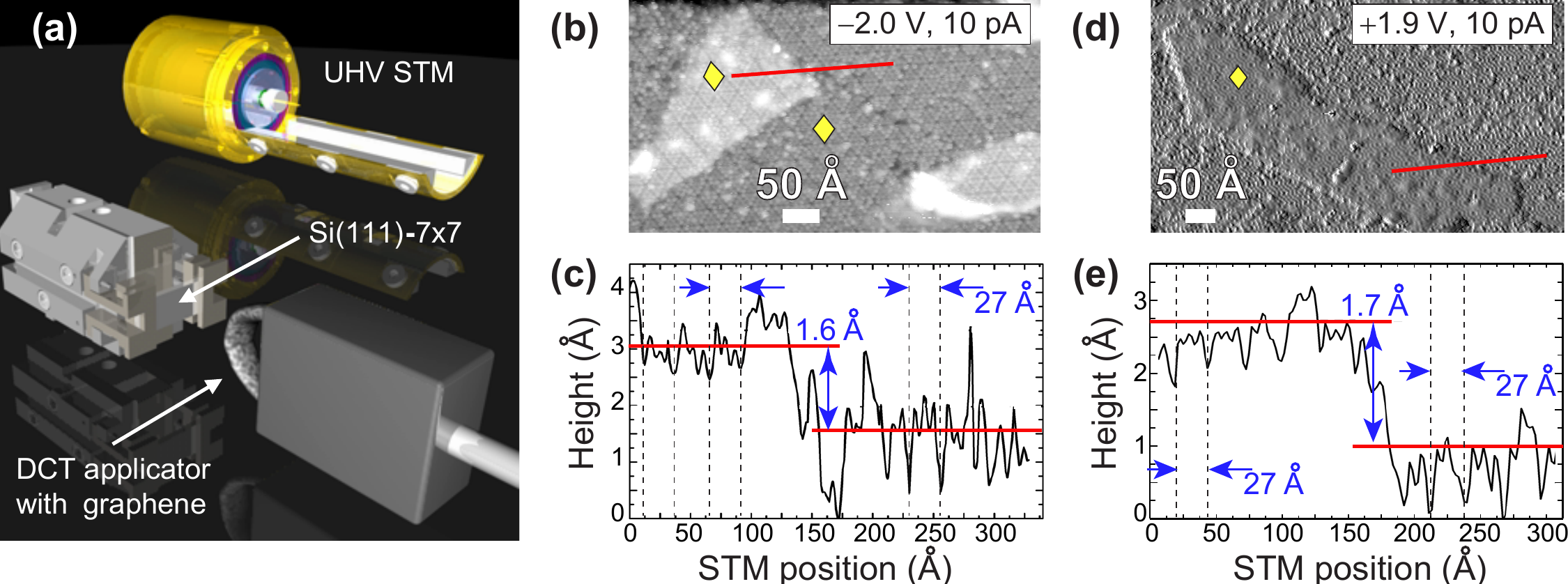}
\caption{(a) Schematics of the dry contact transfer (DCT) technique, whereby a fiberglass applicator transfers graphene to a clean Si(111)$-$7$\times$7 surface {\em in situ}. (b) Filled states ($-$2 V) STM topograph of graphene flakes on the Si(111)-7$\times$7 surface; graphene flakes are seen on a whiter tone. (c) Height profile along the solid line on the STM topograph. The STM height of the graphene feature is $\sim1.6\pm 1.1$ \AA{}. (d) Empty states (+1.9 V) STM derivative image for another graphene flake. (e) Height profile taken along solid line on subplot (d). Diamonds on the STM images indicate the area of a Si(111)$-$7$\times$7 supercell.}
\end{figure*}

\begin{figure}[tb]
\includegraphics[width=0.4\textwidth]{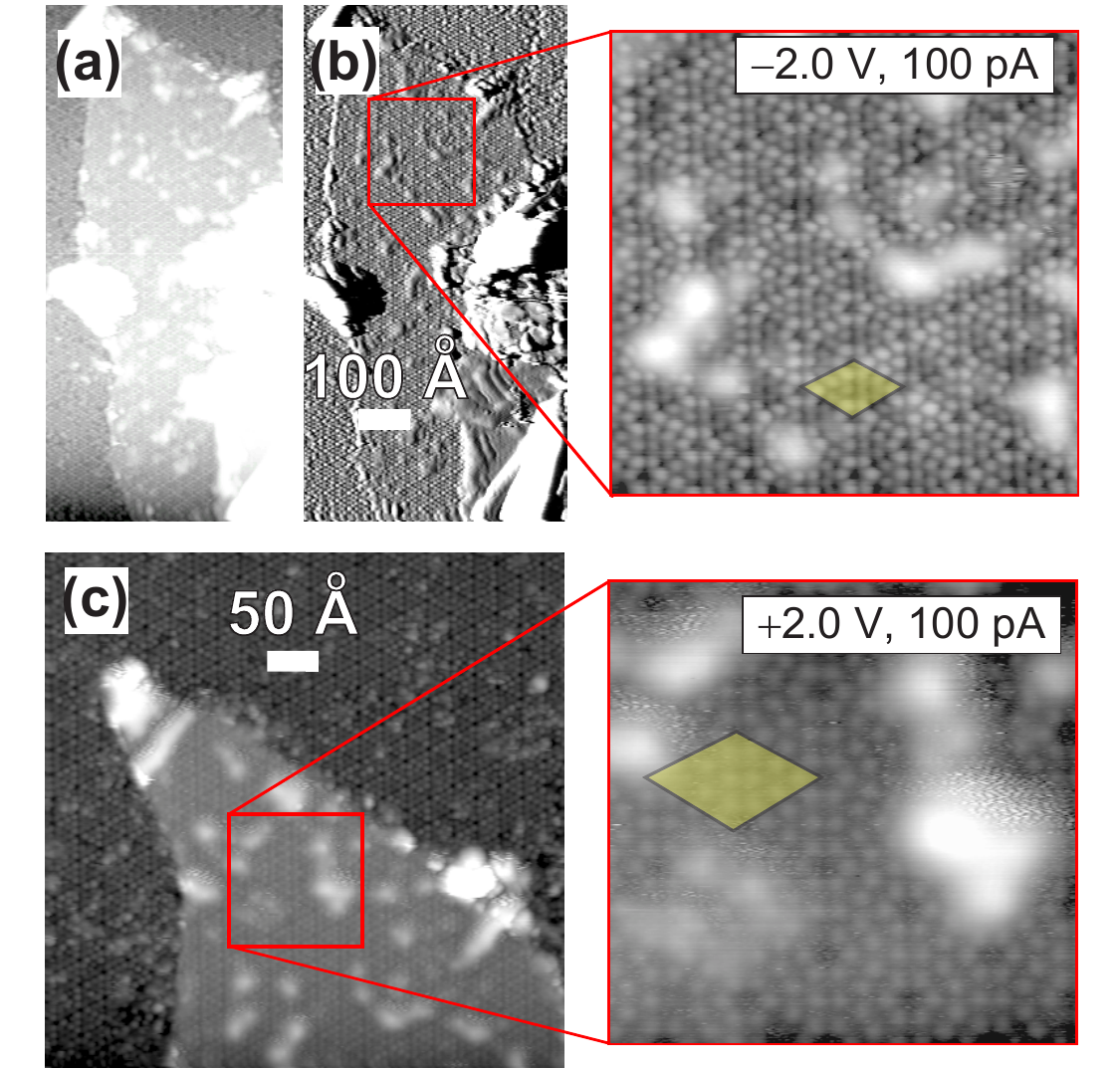}
\caption{(a) Filled-states STM topograph and (b) derivative image of a graphene flake on Si(111). The atomic arrangement of Si(111)$-$7$\times$7 can be clearly seen underneath graphene at the inset. (c) An empty states STM image and its derivative image help establishing that the 7$\times$7 reconstruction of Si(111) remains unchanged after graphene was adsorbed.}
\end{figure}

	To this end, we interface the stable Si(111)$-$7$\times$7 surface with graphene using the dry-contact transfer technique, \cite{Albrecht1} as illustrated in Fig.~1(a). Our transfer process produces a clean interface at the graphene-Si(111) heterostructure, allowing examination with atomically resolved, ultrahigh vacuum (UHV) scanning tunneling microscopy (STM) and spectroscopy (STS). We frame these experimental results and explore the interfacial chemistry by {\em ab-initio} density-functional theory (DFT) calculations using a large supercell in the local density approximation (LDA). These calculations reproduce our experimental findings and reveal the absence of covalent C$-$Si bonds at the graphene-Si(111) interface. Our article complements our previous STM studies of the atomic-scale interaction of carbon-based materials and technologically-relevant silicon substrates.\cite{Albrecht1,Orellana,E1,E2,E3,E4,E5,E6,E8}

In the approach of Fig.~1(a), a fiberglass applicator collects graphene flakes and gently presses them onto a hydrogen-free Si(111) substrate, allowing for subsequent, {\em in situ}, room-temperature STM interrogation. We note that the orientation of graphene flakes from the applicator is random, and, hence, we expect the registry between the graphene and Si(111) lattices to be random also. A filled-states STM image of two graphene flakes on Si(111) is displayed on Fig.~1(b); the graphene flakes are seen on a lighter color on the image. Despite the unknown relative orientation between graphene and Si(111), the diamond-like structure characteristic of Si(111)$-$$7\times 7$ can be seen throughout the image. We superimpose two diamond-like features on the graphene and onto the bare Si(111) surface to help make the area spanned by Si(111) reconstructed supercells more evident. Such apparent registry regardless of relative orientation arises from the fact that the electronic density has a very large contribution from the dangling bonds on Si(111)$-$even in scans acquired over graphene$-$in an effect that we dubbed ``electronic transparency.''\cite{NanoLetters,USPRB2015}

We acquire the height profile of Fig.~1(c) along the line crossing the right side of the flake located on the left side of Fig.~1(b). It possesses an average height difference between the graphene feature and Si(111) of the order of 1.6 \AA. Additionally, the height profile has an oscillatory pattern with a $27$ \AA{} period, highlighted by vertical dashed lines. Si(111) forms a triangular lattice with a lattice constant of 3.81 \AA{}. The $7\times 7$ supercell has, in turn, a period of $7\times 3.81=26.7$ \AA{} that matches the periodicity captured on the height profiles taken on the bare Si(111) substrate and on top of the graphene flake.

\begin{figure}[tb]
\includegraphics[width=0.4\textwidth]{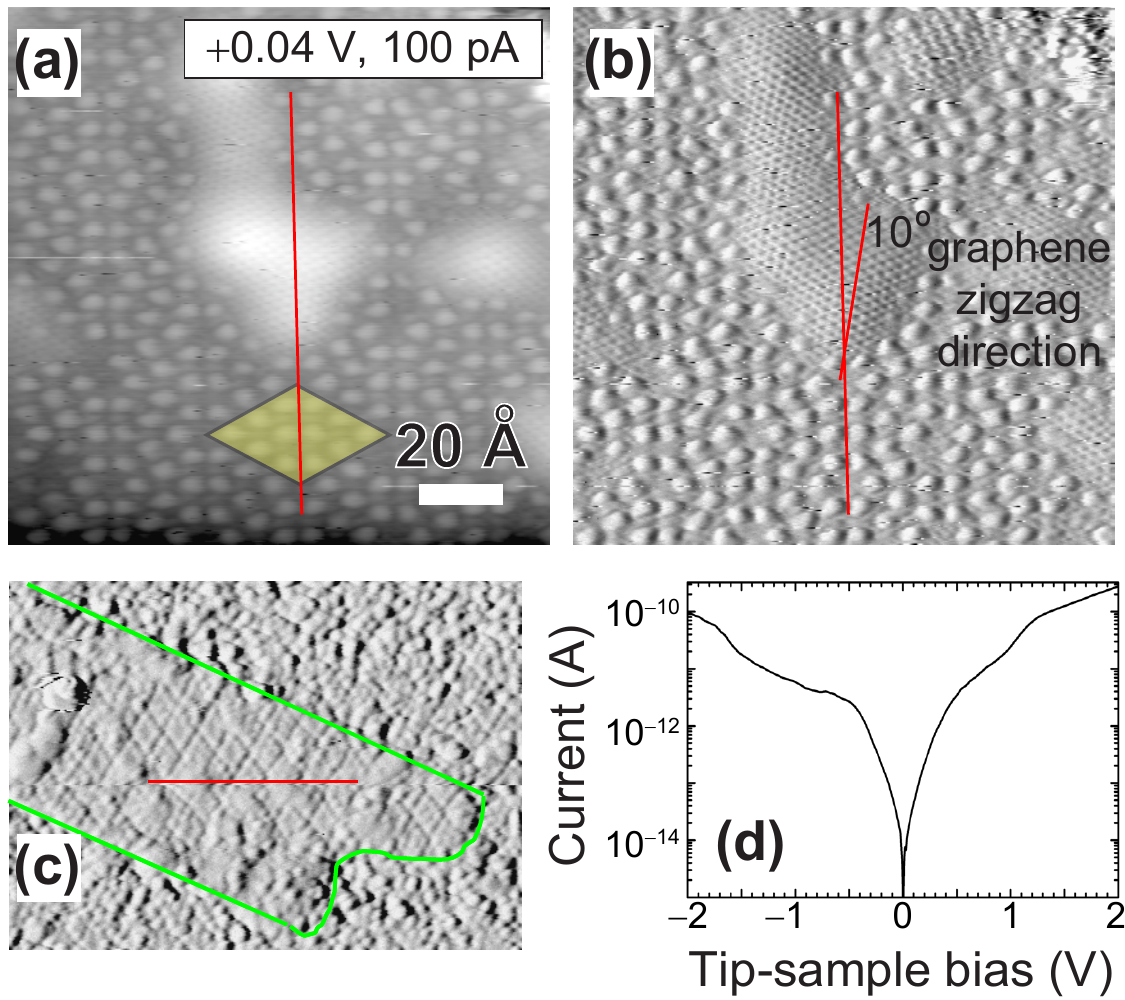}
\caption{(a) STM topograph in empty states of graphene on Si(111). (b) The derivative image shows the relative misalignment among graphene and Si(111)$-$7$\times$7 to be about 10$^o$. (c) Another derivative image of a graphene flake on Si(111)$-$7$\times$7.
(d) Average $I-V$ data over the straight line in (c). No band gap is induced by the substrate or by the narrow width (17.1 nm $\times$ 78 nm) of the flake.}
\end{figure}

For a second graphene flake, we take an empty-states STM derivative image, shown in Fig.~1(d), and its corresponding height profile in Fig.~1(e). Both confirm the oscillatory period of $27$ \AA{} registered in Fig.~1(b). The average STM height for this graphene feature is 1.7 \AA{} higher than the average height of the bare Si(111) substrate, which is similar to the height difference estimated in Fig.~1(c), despite the different relative orientation among graphene and Si(111) for these flakes.

We now inspect subtle effects in the graphene-encapsulated Si(111) substrate by bias-dependent scans. To this end, Fig.~2(a) shows a filled-states STM topographic image of graphene on Si(111), taken at a $-$2 V bias and with a 100 pA tunneling current. Its derivative image is given in Fig.~2(b). These two images permit a clear identification of features characteristic of a Si(111)-$7\times 7$ reconstructed surface. Furthermore, the $7\times 7$ surface reconstruction is seen under empty-states scans, as confirmed in Fig.~2(c). The diamonds drawn on Fig.~2 indicate the unit cell of the silicon surface resolved through the graphene. The results from Fig.~2  help establish that the Si(111)$-$7$\times 7$ surface is not altered chemically or morphologically upon graphene adsorption.

In Fig.~3(a), we discern the relative orientation among graphene and the Si(111)$-$7$\times$7 surface by examining the graphene lattice over bright spots in STM images, where the STM resolves the graphene lattice rather than a convolution with the substrate surface structure. There is a relative misalignment of about 10$^o$ in the zigzag direction of the graphene lattice and the vertical line on the Si(111)$-$7$\times$7 substrate for this particular flake (see derivative image, Fig.~3(b)). The yellow diamond superimposed in Fig.~3(a) reveals features characteristic of Si(111), regardless of the incommensurate moir\`e superlattice between the misaligned graphene and the Si(111)$-$7$\times$7. Thus, even though graphene and Si(111) may have incommensuration and thereby span a large supercell, the Si(111) STM image is not strongly convoluted by the moir\`e superlattice. Consequently, theoretical calculations can reproduce features of the graphene-Si(111) system with the smallest possible supercell, disregarding relative orientation concerns. This simplifying assumption from the experimental data in Figs.~1 to 3 makes comprehensive theoretical analysis possible, as detailed later.

We show a derivative image of another graphene feature in Fig.~3(c), and the straight line in the image corresponds to where we take averaged STS data. Fig. ~3(d) gives these spectra, highlighting features consistent with graphene spectra obtained on other semiconducting substrates by STM.\cite{E5,E6,E8,NanoLetters} Any covalent interactions between the graphene and the Si(111)$-$$7\times 7$ substrate would disrupt the sp$^2$ bonding network, leading to modifcations to the STS data which are absent here.

It is time to discuss the computational study from Ref.~\onlinecite{PRL2013}, and a theory that explains our experimental observations. That work asserts that graphene can be placed on Si(111)$-2\times 1$ and compressed by more than 11\% to establish commensuration. In turn, unidirectional C$-$Si covalent bonds will be created in this extreme compression for graphene on Si(111)$-2\times 1$. These computational results do not describe what we observe and are therefore inconsistent with our findings.

Therefore, we follow our experimental observations which indicate that the features related to silicon dangling bonds are not hampered by a moir\`e superlattice created at the graphene/Si(111)$-7\times 7$ interface. Aiming for correctness and simplicity, we computationally examine the smallest graphene supercell matching Si(111)$-2\times 1$. Fortunately, graphene has a lattice constant $a_{0g}= 2.44$ \AA{}, and a $11\times 11$ graphene supercell matches the Si(111)$-$$7\times 7$ supercell within a negligible mismatch of 0.5 \%.

\begin{figure}[tb]
\includegraphics[width=0.4\textwidth]{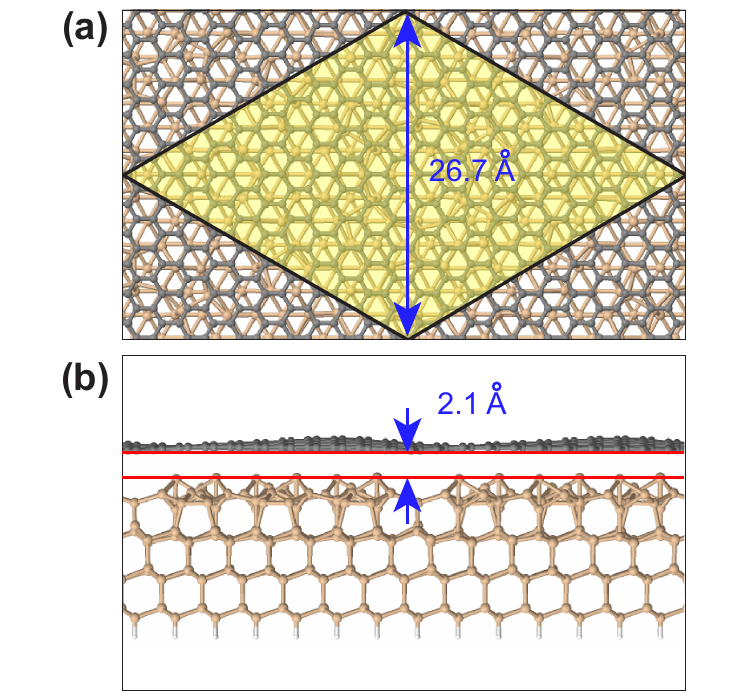}
\caption{(a) Top and (b) side views of a supercell employed in calculations. The closest distance among silicon and carbon atoms is 2.1 \AA{}. The closer distance obtained experimentally ($\sim 1.7$ \AA) may originate from the fact that the STM presses graphene down into the semiconducting substrate to establish the necessary feedback current.}
\end{figure}

\begin{figure}[tb]
\includegraphics[width=0.4\textwidth]{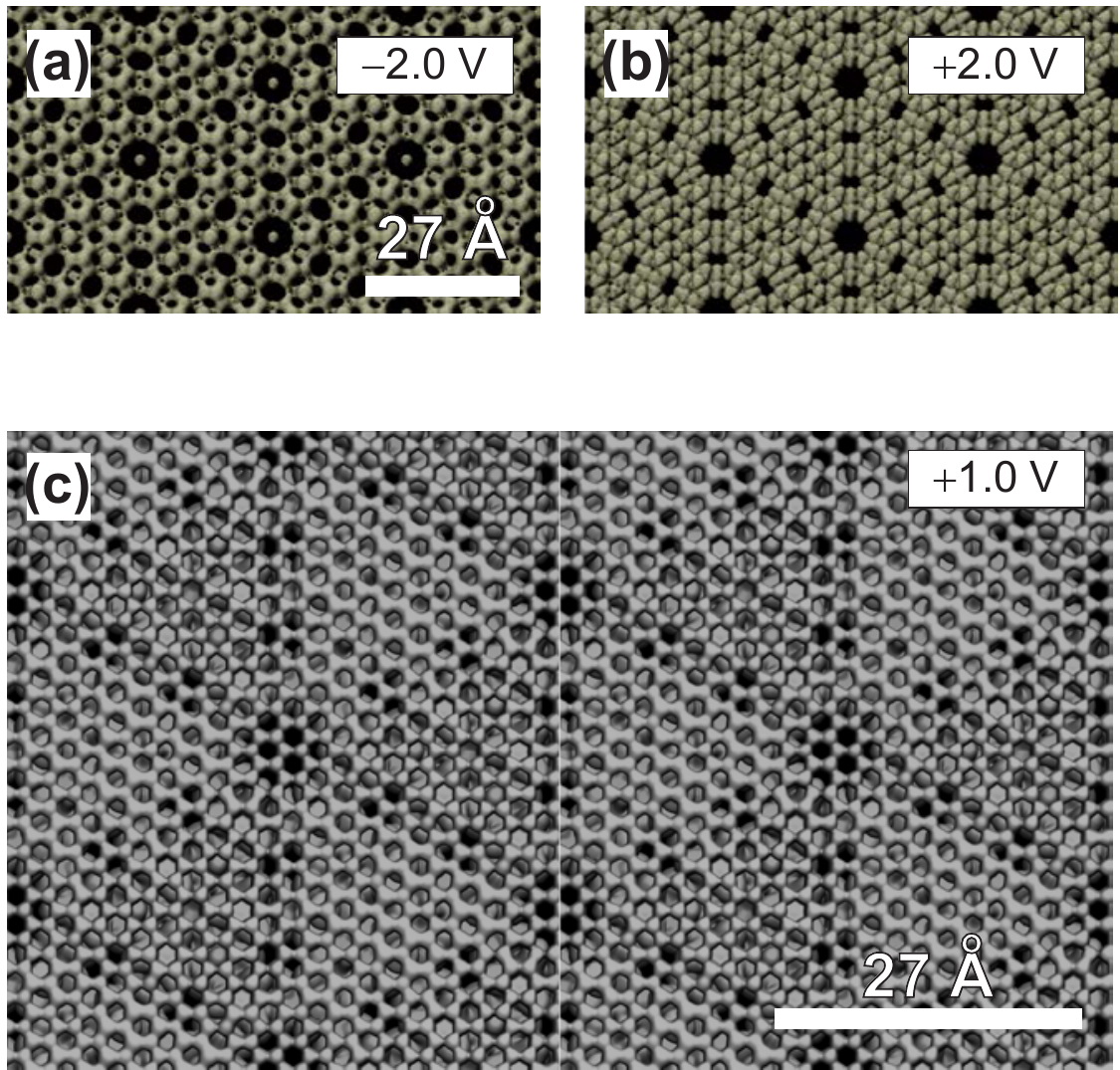}
\caption{(a) Filled and (b) empty-states iso-density surfaces for Si(111)$-$7$\times$7. Comparison of these images to the experimental STM images acquired underneath graphene in Figs.~2(b) and 2(c) helps confirming the lack of atomistic reconstruction on the substrate upon the adsorption of graphene. (c) Empty-states iso-density surface for graphene on Si(111).}
\end{figure}

We create graphene/silicon supercells containing 687 atoms, with a 10 \AA{} vacuum added along the vertical direction separating periodic images. Density-functional theory calculations with the SIESTA code\cite{SIESTA1,SIESTA2} within the local density approximation\cite{LDA1,LDA2} yield reasonable separations among weakly-bonded systems, giving reasonable results at the Si(111)/graphene interface. An equivalent mesh cutoff of 200 Ry is employed, and a conjugate-gradient ionic relaxation is performed until all force components became smaller than 0.02 eV/\AA. The resulting atomistic structure is shown in Fig.~4, and the graphene/Si(111) supercell is emphasized by a yellow diamond. The underlying Si(111) reconstruction is quite similar to the one originally obtained by Brommer and coworkers.\cite{Brommer}

In our calculations, graphene adapts to the Si(111) surface by bending vertically slightly. We note a vertical distance between graphene and the topmost Si atoms that varies between 2.1 and 3.1 \AA{}, for a 1.0 \AA{} variation in graphene height overall. The relative height among weakly bound materials can be further diminished under the dynamical pressure exerted by the STM tip,\cite{NanoLetters} thereby explaining smaller heights seen experimentally.

Top views of the simulated density isosurfaces for Si(111)$-$7$\times$7 are displayed in Figs.~5(a) and 5(b) and for the equilibrium structure containing graphene (c.f., Fig.~4) in Fig.~5(c). A clear superposition of graphene and Si(111) features is apparent in Fig.~5(c). The electronic hybridization among these systems increases when graphene is pushed in closer proximity to Si(111), revealing the semiconducting substrate's features more clearly, consistent with our experimental results. Our computational results show no covalent C$-$Si bonds between the graphene and Si(111) in the equilibrium structure and, more importantly, they reproduce the periodicity seen in the experimental STM images.

In summary, we studied monolayer graphene flakes adsorbed on the Si(111)$-$7$\times$7 substrate with a combination of STM measurements and {\em ab-initio} calculations. The characteristic 7$\times$7 reconstruction of this semiconductor substrate was resolved through graphene in all of our scans, so its atomistic configuration was not altered upon the adsorption of graphene. No chemical bonding among graphene and silicon atoms occurred within this system. Therefore, we demonstrated that the typical $7\times 7$ reconstruction of Si(111) remains unchanged after graphene adsorption on this highly reactive surface. Our report demonstrates graphene's use as a protective layer for inherently reactive surfaces like Si(111), facilitating {\em ex-situ} handling of these heterostructures. Additionally, these results help ongoing efforts in the integration of graphene with silicon for electronic and optoelectronic applications.

We acknowledge funding from the Office of Naval Research (ONR) through grants N00014-06-10120 and N00014-09-0180, the National Defense Science and Engineering Graduate Fellowship (NDSEG) through the Army Research Office (ARO), and XSEDE (TG-PHY090002) and Arkansas' Razor II for computer resources.

\nocite{*}
%\bibliography{manuscript}% Produces the bibliography via BibTeX.

\begin{thebibliography}{41}%
\makeatletter
\providecommand \@ifxundefined [1]{%
 \@ifx{#1\undefined}
}%
\providecommand \@ifnum [1]{%
 \ifnum #1\expandafter \@firstoftwo
 \else \expandafter \@secondoftwo
 \fi
}%
\providecommand \@ifx [1]{%
 \ifx #1\expandafter \@firstoftwo
 \else \expandafter \@secondoftwo
 \fi
}%
\providecommand \natexlab [1]{#1}%
\providecommand \enquote  [1]{``#1''}%
\providecommand \bibnamefont  [1]{#1}%
\providecommand \bibfnamefont [1]{#1}%
\providecommand \citenamefont [1]{#1}%
\providecommand \href@noop [0]{\@secondoftwo}%
\providecommand \href [0]{\begingroup \@sanitize@url \@href}%
\providecommand \@href[1]{\@@startlink{#1}\@@href}%
\providecommand \@@href[1]{\endgroup#1\@@endlink}%
\providecommand \@sanitize@url [0]{\catcode `\\12\catcode `\$12\catcode
  `\&12\catcode `\#12\catcode `\^12\catcode `\_12\catcode `\%12\relax}%
\providecommand \@@startlink[1]{}%
\providecommand \@@endlink[0]{}%
\providecommand \url  [0]{\begingroup\@sanitize@url \@url }%
\providecommand \@url [1]{\endgroup\@href {#1}{\urlprefix }}%
\providecommand \urlprefix  [0]{URL }%
\providecommand \Eprint [0]{\href }%
\providecommand \doibase [0]{http://dx.doi.org/}%
\providecommand \selectlanguage [0]{\@gobble}%
\providecommand \bibinfo  [0]{\@secondoftwo}%
\providecommand \bibfield  [0]{\@secondoftwo}%
\providecommand \translation [1]{[#1]}%
\providecommand \BibitemOpen [0]{}%
\providecommand \bibitemStop [0]{}%
\providecommand \bibitemNoStop [0]{.\EOS\space}%
\providecommand \EOS [0]{\spacefactor3000\relax}%
\providecommand \BibitemShut  [1]{\csname bibitem#1\endcsname}%
\let\auto@bib@innerbib\@empty
%</preamble>
\bibitem [{\citenamefont {Neto}\ \emph {et~al.}(2009)\citenamefont {Neto},
  \citenamefont {Guinea}, \citenamefont {Peres}, \citenamefont {Novoselov},\
  and\ \citenamefont {Geim}}]{RMP}%
  \BibitemOpen
  \bibfield  {author} {\bibinfo {author} {\bibfnamefont {A.~H.~C.}\
  \bibnamefont {Neto}}, \bibinfo {author} {\bibfnamefont {F.}~\bibnamefont
  {Guinea}}, \bibinfo {author} {\bibfnamefont {N.~M.~R.}\ \bibnamefont
  {Peres}}, \bibinfo {author} {\bibfnamefont {K.~S.}\ \bibnamefont
  {Novoselov}}, \ and\ \bibinfo {author} {\bibfnamefont {A.~K.}\ \bibnamefont
  {Geim}},\ }\href@noop {} {\bibfield  {journal} {\bibinfo  {journal} {Rev.
  Mod. Phys.}\ }\textbf {\bibinfo {volume} {81}},\ \bibinfo {pages} {109}
  (\bibinfo {year} {2009})}\BibitemShut {NoStop}%
\bibitem [{\citenamefont {Katsnelson}(2012)}]{KatsnelsonBook}%
  \BibitemOpen
  \bibfield  {author} {\bibinfo {author} {\bibfnamefont {M.~I.}\ \bibnamefont
  {Katsnelson}},\ }\href@noop {} {\emph {\bibinfo {title} {Graphene: Carbon in
  two dimensions}}},\ \bibinfo {edition} {1st}\ ed.\ (\bibinfo  {publisher}
  {Cambdridge U. Press},\ \bibinfo {address} {Cambridge},\ \bibinfo {year}
  {2012})\BibitemShut {NoStop}%
\bibitem [{\citenamefont {Bolotin}\ \emph {et~al.}(2008)\citenamefont
  {Bolotin}, \citenamefont {Sikes}, \citenamefont {Jiang}, \citenamefont
  {Klima}, \citenamefont {Fudenberg}, \citenamefont {Hone}, \citenamefont
  {Kim},\ and\ \citenamefont {Stormer}}]{r1}%
  \BibitemOpen
  \bibfield  {author} {\bibinfo {author} {\bibfnamefont {K.~I.}\ \bibnamefont
  {Bolotin}}, \bibinfo {author} {\bibfnamefont {K.~J.}\ \bibnamefont {Sikes}},
  \bibinfo {author} {\bibfnamefont {Z.}~\bibnamefont {Jiang}}, \bibinfo
  {author} {\bibfnamefont {M.}~\bibnamefont {Klima}}, \bibinfo {author}
  {\bibfnamefont {G.}~\bibnamefont {Fudenberg}}, \bibinfo {author}
  {\bibfnamefont {J.}~\bibnamefont {Hone}}, \bibinfo {author} {\bibfnamefont
  {P.}~\bibnamefont {Kim}}, \ and\ \bibinfo {author} {\bibfnamefont {H.~L.}\
  \bibnamefont {Stormer}},\ }\href@noop {} {\bibfield  {journal} {\bibinfo
  {journal} {Solid State Comm.}\ }\textbf {\bibinfo {volume} {146}},\ \bibinfo
  {pages} {351} (\bibinfo {year} {2008})}\BibitemShut {NoStop}%
\bibitem [{\citenamefont {Sarma}\ \emph {et~al.}(2011)\citenamefont {Sarma},
  \citenamefont {Adam}, \citenamefont {Hwang},\ and\ \citenamefont
  {Rossi}}]{r2}%
  \BibitemOpen
  \bibfield  {author} {\bibinfo {author} {\bibfnamefont {S.~D.}\ \bibnamefont
  {Sarma}}, \bibinfo {author} {\bibfnamefont {S.}~\bibnamefont {Adam}},
  \bibinfo {author} {\bibfnamefont {E.~H.}\ \bibnamefont {Hwang}}, \ and\
  \bibinfo {author} {\bibfnamefont {E.}~\bibnamefont {Rossi}},\ }\href@noop {}
  {\bibfield  {journal} {\bibinfo  {journal} {Rev. Mod. Phys.}\ }\textbf
  {\bibinfo {volume} {83}},\ \bibinfo {pages} {407} (\bibinfo {year}
  {2011})}\BibitemShut {NoStop}%
\bibitem [{\citenamefont {Balandin}\ \emph {et~al.}(2008)\citenamefont
  {Balandin}, \citenamefont {Ghosh}, \citenamefont {Bao}, \citenamefont
  {Calizo}, \citenamefont {Teweldebrhan}, \citenamefont {Miao},\ and\
  \citenamefont {Lau}}]{r3}%
  \BibitemOpen
  \bibfield  {author} {\bibinfo {author} {\bibfnamefont {A.~A.}\ \bibnamefont
  {Balandin}}, \bibinfo {author} {\bibfnamefont {S.}~\bibnamefont {Ghosh}},
  \bibinfo {author} {\bibfnamefont {W.}~\bibnamefont {Bao}}, \bibinfo {author}
  {\bibfnamefont {I.}~\bibnamefont {Calizo}}, \bibinfo {author} {\bibfnamefont
  {D.}~\bibnamefont {Teweldebrhan}}, \bibinfo {author} {\bibfnamefont
  {F.}~\bibnamefont {Miao}}, \ and\ \bibinfo {author} {\bibfnamefont {C.~N.}\
  \bibnamefont {Lau}},\ }\href@noop {} {\bibfield  {journal} {\bibinfo
  {journal} {Nano Lett.}\ }\textbf {\bibinfo {volume} {8}},\ \bibinfo {pages}
  {902} (\bibinfo {year} {2008})}\BibitemShut {NoStop}%
\bibitem [{\citenamefont {Binnig}\ \emph {et~al.}(1983)\citenamefont {Binnig},
  \citenamefont {Rohrer}, \citenamefont {Gerber},\ and\ \citenamefont
  {Weibel}}]{Binnig}%
  \BibitemOpen
  \bibfield  {author} {\bibinfo {author} {\bibfnamefont {G.}~\bibnamefont
  {Binnig}}, \bibinfo {author} {\bibfnamefont {H.}~\bibnamefont {Rohrer}},
  \bibinfo {author} {\bibfnamefont {C.}~\bibnamefont {Gerber}}, \ and\ \bibinfo
  {author} {\bibfnamefont {E.}~\bibnamefont {Weibel}},\ }\href@noop {}
  {\bibfield  {journal} {\bibinfo  {journal} {Phys. Rev. Lett.}\ }\textbf
  {\bibinfo {volume} {50}},\ \bibinfo {pages} {120} (\bibinfo {year}
  {1983})}\BibitemShut {NoStop}%
\bibitem [{\citenamefont {Himpsel}\ and\ \citenamefont
  {Batra}(1984)}]{Himpsel}%
  \BibitemOpen
  \bibfield  {author} {\bibinfo {author} {\bibfnamefont {F.}~\bibnamefont
  {Himpsel}}\ and\ \bibinfo {author} {\bibfnamefont {I.}~\bibnamefont
  {Batra}},\ }\href@noop {} {\bibfield  {journal} {\bibinfo  {journal} {J. Vac.
  Sci. Technol. A}\ }\textbf {\bibinfo {volume} {2}},\ \bibinfo {pages} {952}
  (\bibinfo {year} {1984})}\BibitemShut {NoStop}%
\bibitem [{\citenamefont {Takayanagi}\ \emph {et~al.}(1985)\citenamefont
  {Takayanagi}, \citenamefont {Tanishiro}, \citenamefont {Takahashi},\ and\
  \citenamefont {Takahashi}}]{Takayanagi}%
  \BibitemOpen
  \bibfield  {author} {\bibinfo {author} {\bibfnamefont {K.}~\bibnamefont
  {Takayanagi}}, \bibinfo {author} {\bibfnamefont {Y.}~\bibnamefont
  {Tanishiro}}, \bibinfo {author} {\bibfnamefont {M.}~\bibnamefont
  {Takahashi}}, \ and\ \bibinfo {author} {\bibfnamefont {S.}~\bibnamefont
  {Takahashi}},\ }\href@noop {} {\bibfield  {journal} {\bibinfo  {journal} {J.
  Vac. Sci. Technol. A}\ }\textbf {\bibinfo {volume} {3}},\ \bibinfo {pages}
  {1502} (\bibinfo {year} {1985})}\BibitemShut {NoStop}%
\bibitem [{\citenamefont {Brommer}\ \emph {et~al.}(1992)\citenamefont
  {Brommer}, \citenamefont {Needels}, \citenamefont {Larson},\ and\
  \citenamefont {Joannopoulos}}]{Brommer}%
  \BibitemOpen
  \bibfield  {author} {\bibinfo {author} {\bibfnamefont {K.}~\bibnamefont
  {Brommer}}, \bibinfo {author} {\bibfnamefont {M.}~\bibnamefont {Needels}},
  \bibinfo {author} {\bibfnamefont {B.}~\bibnamefont {Larson}}, \ and\ \bibinfo
  {author} {\bibfnamefont {J.}~\bibnamefont {Joannopoulos}},\ }\href@noop {}
  {\bibfield  {journal} {\bibinfo  {journal} {Phys. Rev. Lett.}\ }\textbf
  {\bibinfo {volume} {68}},\ \bibinfo {pages} {1355} (\bibinfo {year}
  {1992})}\BibitemShut {NoStop}%
\bibitem [{\citenamefont {Pandey}(1981)}]{Pandey}%
  \BibitemOpen
  \bibfield  {author} {\bibinfo {author} {\bibfnamefont {K.}~\bibnamefont
  {Pandey}},\ }\href@noop {} {\bibfield  {journal} {\bibinfo  {journal} {Phys.
  Rev. Lett.}\ }\textbf {\bibinfo {volume} {47}},\ \bibinfo {pages} {1913}
  (\bibinfo {year} {1981})}\BibitemShut {NoStop}%
\bibitem [{\citenamefont {Himpsel}\ \emph {et~al.}(1984)\citenamefont
  {Himpsel}, \citenamefont {Marcus}, \citenamefont {Tromp}, \citenamefont
  {Batra}, \citenamefont {Cook}, \citenamefont {Jona},\ and\ \citenamefont
  {Liu}}]{Himpsel2}%
  \BibitemOpen
  \bibfield  {author} {\bibinfo {author} {\bibfnamefont {F.}~\bibnamefont
  {Himpsel}}, \bibinfo {author} {\bibfnamefont {P.}~\bibnamefont {Marcus}},
  \bibinfo {author} {\bibfnamefont {R.}~\bibnamefont {Tromp}}, \bibinfo
  {author} {\bibfnamefont {I.}~\bibnamefont {Batra}}, \bibinfo {author}
  {\bibfnamefont {M.}~\bibnamefont {Cook}}, \bibinfo {author} {\bibfnamefont
  {F.}~\bibnamefont {Jona}}, \ and\ \bibinfo {author} {\bibfnamefont
  {H.}~\bibnamefont {Liu}},\ }\href@noop {} {\bibfield  {journal} {\bibinfo
  {journal} {Phys. Rev. B}\ }\textbf {\bibinfo {volume} {30}},\ \bibinfo
  {pages} {2257} (\bibinfo {year} {1984})}\BibitemShut {NoStop}%
\bibitem [{\citenamefont {Stroscio}, \citenamefont {Feenstra},\ and\
  \citenamefont {Fein}(1986)}]{Stroscio}%
  \BibitemOpen
  \bibfield  {author} {\bibinfo {author} {\bibfnamefont {J.}~\bibnamefont
  {Stroscio}}, \bibinfo {author} {\bibfnamefont {R.}~\bibnamefont {Feenstra}},
  \ and\ \bibinfo {author} {\bibfnamefont {A.}~\bibnamefont {Fein}},\
  }\href@noop {} {\bibfield  {journal} {\bibinfo  {journal} {Phys. Rev. Lett.}\
  }\textbf {\bibinfo {volume} {57}},\ \bibinfo {pages} {2579} (\bibinfo {year}
  {1986})}\BibitemShut {NoStop}%
\bibitem [{\citenamefont {Ancilotto}\ \emph {et~al.}(1990)\citenamefont
  {Ancilotto}, \citenamefont {Andreoni}, \citenamefont {Selloni}, \citenamefont
  {Car},\ and\ \citenamefont {Parinello}}]{Ancilotto}%
  \BibitemOpen
  \bibfield  {author} {\bibinfo {author} {\bibfnamefont {F.}~\bibnamefont
  {Ancilotto}}, \bibinfo {author} {\bibfnamefont {W.}~\bibnamefont {Andreoni}},
  \bibinfo {author} {\bibfnamefont {A.}~\bibnamefont {Selloni}}, \bibinfo
  {author} {\bibfnamefont {R.}~\bibnamefont {Car}}, \ and\ \bibinfo {author}
  {\bibfnamefont {M.}~\bibnamefont {Parinello}},\ }\href@noop {} {\bibfield
  {journal} {\bibinfo  {journal} {Phys. Rev. Lett.}\ }\textbf {\bibinfo
  {volume} {65}},\ \bibinfo {pages} {3148} (\bibinfo {year}
  {1990})}\BibitemShut {NoStop}%
\bibitem [{\citenamefont {Spence}(1988)}]{Spence}%
  \BibitemOpen
  \bibfield  {author} {\bibinfo {author} {\bibfnamefont {J.}~\bibnamefont
  {Spence}},\ }\href@noop {} {\bibfield  {journal} {\bibinfo  {journal} {Phys.
  Rev. B}\ }\textbf {\bibinfo {volume} {38}},\ \bibinfo {pages} {12672}
  (\bibinfo {year} {1988})}\BibitemShut {NoStop}%
\bibitem [{\citenamefont {Ritter}\ and\ \citenamefont {Lyding}(2008)}]{E5}%
  \BibitemOpen
  \bibfield  {author} {\bibinfo {author} {\bibfnamefont {K.~A.}\ \bibnamefont
  {Ritter}}\ and\ \bibinfo {author} {\bibfnamefont {J.~W.}\ \bibnamefont
  {Lyding}},\ }\href@noop {} {\bibfield  {journal} {\bibinfo  {journal}
  {Nanotechnology}\ }\textbf {\bibinfo {volume} {19}},\ \bibinfo {pages}
  {015704} (\bibinfo {year} {2008})}\BibitemShut {NoStop}%
\bibitem [{\citenamefont {Ritter}\ and\ \citenamefont {Lyding}(2009)}]{E6}%
  \BibitemOpen
  \bibfield  {author} {\bibinfo {author} {\bibfnamefont {K.~A.}\ \bibnamefont
  {Ritter}}\ and\ \bibinfo {author} {\bibfnamefont {J.~W.}\ \bibnamefont
  {Lyding}},\ }\href@noop {} {\bibfield  {journal} {\bibinfo  {journal} {Nature
  Matter.}\ }\textbf {\bibinfo {volume} {8}},\ \bibinfo {pages} {235} (\bibinfo
  {year} {2009})}\BibitemShut {NoStop}%
\bibitem [{\citenamefont {Xu}\ \emph {et~al.}(2011)\citenamefont {Xu},
  \citenamefont {He}, \citenamefont {Schmucker}, \citenamefont {Guo},
  \citenamefont {Wood}, \citenamefont {Koepke}, \citenamefont {Lyding},\ and\
  \citenamefont {Aluru}}]{E8}%
  \BibitemOpen
  \bibfield  {author} {\bibinfo {author} {\bibfnamefont {Y.}~\bibnamefont
  {Xu}}, \bibinfo {author} {\bibfnamefont {K.~T.}\ \bibnamefont {He}}, \bibinfo
  {author} {\bibfnamefont {S.~W.}\ \bibnamefont {Schmucker}}, \bibinfo {author}
  {\bibfnamefont {Z.}~\bibnamefont {Guo}}, \bibinfo {author} {\bibfnamefont
  {J.~D.}\ \bibnamefont {Wood}}, \bibinfo {author} {\bibfnamefont {J.~C.}\
  \bibnamefont {Koepke}}, \bibinfo {author} {\bibfnamefont {J.~W.}\
  \bibnamefont {Lyding}}, \ and\ \bibinfo {author} {\bibfnamefont {N.~R.}\
  \bibnamefont {Aluru}},\ }\href@noop {} {\bibfield  {journal} {\bibinfo
  {journal} {Nano Lett.}\ }\textbf {\bibinfo {volume} {11}},\ \bibinfo {pages}
  {2735} (\bibinfo {year} {2011})}\BibitemShut {NoStop}%
\bibitem [{\citenamefont {He}\ \emph {et~al.}(2010)\citenamefont {He},
  \citenamefont {Koepke}, \citenamefont {Barraza-Lopez},\ and\ \citenamefont
  {Lyding}}]{NanoLetters}%
  \BibitemOpen
  \bibfield  {author} {\bibinfo {author} {\bibfnamefont {K.~T.}\ \bibnamefont
  {He}}, \bibinfo {author} {\bibfnamefont {J.~C.}\ \bibnamefont {Koepke}},
  \bibinfo {author} {\bibfnamefont {S.}~\bibnamefont {Barraza-Lopez}}, \ and\
  \bibinfo {author} {\bibfnamefont {J.~W.}\ \bibnamefont {Lyding}},\
  }\href@noop {} {\bibfield  {journal} {\bibinfo  {journal} {Nano Lett.}\
  }\textbf {\bibinfo {volume} {10}},\ \bibinfo {pages} {3446} (\bibinfo {year}
  {2010})}\BibitemShut {NoStop}%
\bibitem [{\citenamefont {Sprinkle}\ \emph {et~al.}(2010)\citenamefont
  {Sprinkle}, \citenamefont {Ruan}, \citenamefont {Hu}, \citenamefont
  {Hankinson}, \citenamefont {Rubio-Roy}, \citenamefont {Zhang}, \citenamefont
  {Wu}, \citenamefont {Berger},\ and\ \citenamefont {de~Heer}}]{SC1}%
  \BibitemOpen
  \bibfield  {author} {\bibinfo {author} {\bibfnamefont {M.}~\bibnamefont
  {Sprinkle}}, \bibinfo {author} {\bibfnamefont {M.}~\bibnamefont {Ruan}},
  \bibinfo {author} {\bibfnamefont {Y.}~\bibnamefont {Hu}}, \bibinfo {author}
  {\bibfnamefont {J.}~\bibnamefont {Hankinson}}, \bibinfo {author}
  {\bibfnamefont {M.}~\bibnamefont {Rubio-Roy}}, \bibinfo {author}
  {\bibfnamefont {B.}~\bibnamefont {Zhang}}, \bibinfo {author} {\bibfnamefont
  {X.}~\bibnamefont {Wu}}, \bibinfo {author} {\bibfnamefont {C.}~\bibnamefont
  {Berger}}, \ and\ \bibinfo {author} {\bibfnamefont {W.~A.}\ \bibnamefont
  {de~Heer}},\ }\href@noop {} {\bibfield  {journal} {\bibinfo  {journal}
  {Nature Nanotech.}\ }\textbf {\bibinfo {volume} {5}},\ \bibinfo {pages} {727}
  (\bibinfo {year} {2010})}\BibitemShut {NoStop}%
\bibitem [{\citenamefont {Rutter}\ \emph
  {et~al.}(2007{\natexlab{a}})\citenamefont {Rutter}, \citenamefont
  {Guisinger}, \citenamefont {Crain}, \citenamefont {Jarvis}, \citenamefont
  {Stiles}, \citenamefont {Li}, \citenamefont {First},\ and\ \citenamefont
  {Stroscio}}]{SC2}%
  \BibitemOpen
  \bibfield  {author} {\bibinfo {author} {\bibfnamefont {G.~M.}\ \bibnamefont
  {Rutter}}, \bibinfo {author} {\bibfnamefont {N.~P.}\ \bibnamefont
  {Guisinger}}, \bibinfo {author} {\bibfnamefont {J.~N.}\ \bibnamefont
  {Crain}}, \bibinfo {author} {\bibfnamefont {E.~A.~A.}\ \bibnamefont
  {Jarvis}}, \bibinfo {author} {\bibfnamefont {M.~D.}\ \bibnamefont {Stiles}},
  \bibinfo {author} {\bibfnamefont {T.}~\bibnamefont {Li}}, \bibinfo {author}
  {\bibfnamefont {P.~N.}\ \bibnamefont {First}}, \ and\ \bibinfo {author}
  {\bibfnamefont {J.~A.}\ \bibnamefont {Stroscio}},\ }\href@noop {} {\bibfield
  {journal} {\bibinfo  {journal} {Phys. Rev. B}\ }\textbf {\bibinfo {volume}
  {76}},\ \bibinfo {pages} {235416} (\bibinfo {year}
  {2007}{\natexlab{a}})}\BibitemShut {NoStop}%
\bibitem [{\citenamefont {Rutter}\ \emph
  {et~al.}(2007{\natexlab{b}})\citenamefont {Rutter}, \citenamefont {Crain},
  \citenamefont {Guisinger}, \citenamefont {Li}, \citenamefont {First},\ and\
  \citenamefont {Stroscio}}]{SC3}%
  \BibitemOpen
  \bibfield  {author} {\bibinfo {author} {\bibfnamefont {G.~M.}\ \bibnamefont
  {Rutter}}, \bibinfo {author} {\bibfnamefont {J.~N.}\ \bibnamefont {Crain}},
  \bibinfo {author} {\bibfnamefont {N.~P.}\ \bibnamefont {Guisinger}}, \bibinfo
  {author} {\bibfnamefont {T.}~\bibnamefont {Li}}, \bibinfo {author}
  {\bibfnamefont {P.~N.}\ \bibnamefont {First}}, \ and\ \bibinfo {author}
  {\bibfnamefont {J.~A.}\ \bibnamefont {Stroscio}},\ }\href@noop {} {\bibfield
  {journal} {\bibinfo  {journal} {Science}\ }\textbf {\bibinfo {volume}
  {317}},\ \bibinfo {pages} {219} (\bibinfo {year}
  {2007}{\natexlab{b}})}\BibitemShut {NoStop}%
\bibitem [{\citenamefont {Kim}\ \emph {et~al.}(2008)\citenamefont {Kim},
  \citenamefont {Ihm}, \citenamefont {Choi},\ and\ \citenamefont
  {Son}}]{Korean}%
  \BibitemOpen
  \bibfield  {author} {\bibinfo {author} {\bibfnamefont {S.}~\bibnamefont
  {Kim}}, \bibinfo {author} {\bibfnamefont {J.}~\bibnamefont {Ihm}}, \bibinfo
  {author} {\bibfnamefont {H.~J.}\ \bibnamefont {Choi}}, \ and\ \bibinfo
  {author} {\bibfnamefont {Y.-W.}\ \bibnamefont {Son}},\ }\href@noop {}
  {\bibfield  {journal} {\bibinfo  {journal} {Phys. Rev. Lett.}\ }\textbf
  {\bibinfo {volume} {100}},\ \bibinfo {pages} {176802} (\bibinfo {year}
  {2008})}\BibitemShut {NoStop}%
\bibitem [{\citenamefont {Giovannetti}\ \emph {et~al.}(2007)\citenamefont
  {Giovannetti}, \citenamefont {Khomyakov}, \citenamefont {Brocks},
  \citenamefont {Kelly},\ and\ \citenamefont {van~den Brink}}]{hbnPRL}%
  \BibitemOpen
  \bibfield  {author} {\bibinfo {author} {\bibfnamefont {G.}~\bibnamefont
  {Giovannetti}}, \bibinfo {author} {\bibfnamefont {P.~A.}\ \bibnamefont
  {Khomyakov}}, \bibinfo {author} {\bibfnamefont {G.}~\bibnamefont {Brocks}},
  \bibinfo {author} {\bibfnamefont {P.~J.}\ \bibnamefont {Kelly}}, \ and\
  \bibinfo {author} {\bibfnamefont {J.}~\bibnamefont {van~den Brink}},\
  }\href@noop {} {\bibfield  {journal} {\bibinfo  {journal} {Phys. Rev. B}\
  }\textbf {\bibinfo {volume} {76}},\ \bibinfo {pages} {073103} (\bibinfo
  {year} {2007})}\BibitemShut {NoStop}%
\bibitem [{\citenamefont {Xue}\ \emph {et~al.}(2011)\citenamefont {Xue},
  \citenamefont {Sanchez-Yamagishi}, \citenamefont {Bulmash}, \citenamefont
  {Jacquod}, \citenamefont {Deshpande}, \citenamefont {Watanabe}, \citenamefont
  {T.~Taniguchi},\ and\ \citenamefont {LeRoy}}]{hbnLeRoy}%
  \BibitemOpen
  \bibfield  {author} {\bibinfo {author} {\bibfnamefont {J.}~\bibnamefont
  {Xue}}, \bibinfo {author} {\bibfnamefont {J.}~\bibnamefont
  {Sanchez-Yamagishi}}, \bibinfo {author} {\bibfnamefont {D.}~\bibnamefont
  {Bulmash}}, \bibinfo {author} {\bibfnamefont {P.}~\bibnamefont {Jacquod}},
  \bibinfo {author} {\bibfnamefont {A.}~\bibnamefont {Deshpande}}, \bibinfo
  {author} {\bibfnamefont {K.}~\bibnamefont {Watanabe}}, \bibinfo {author}
  {\bibfnamefont {a.~J.-H.}\ \bibnamefont {T.~Taniguchi}}, \ and\ \bibinfo
  {author} {\bibfnamefont {B.~J.}\ \bibnamefont {LeRoy}},\ }\href@noop {}
  {\bibfield  {journal} {\bibinfo  {journal} {Nature Materials}\ }\textbf
  {\bibinfo {volume} {10}},\ \bibinfo {pages} {282} (\bibinfo {year}
  {2011})}\BibitemShut {NoStop}%
\bibitem [{\citenamefont {Kindermann}, \citenamefont {Uchoa},\ and\
  \citenamefont {Miller}(2012)}]{hbn1}%
  \BibitemOpen
  \bibfield  {author} {\bibinfo {author} {\bibfnamefont {M.}~\bibnamefont
  {Kindermann}}, \bibinfo {author} {\bibfnamefont {B.}~\bibnamefont {Uchoa}}, \
  and\ \bibinfo {author} {\bibfnamefont {D.~L.}\ \bibnamefont {Miller}},\
  }\href@noop {} {\bibfield  {journal} {\bibinfo  {journal} {Phys. Rev. B}\
  }\textbf {\bibinfo {volume} {86}},\ \bibinfo {pages} {115415} (\bibinfo
  {year} {2012})}\BibitemShut {NoStop}%
\bibitem [{\citenamefont {Amet}\ \emph {et~al.}(2013)\citenamefont {Amet},
  \citenamefont {Williams}, \citenamefont {Watanabe}, \citenamefont
  {Taniguchi},\ and\ \citenamefont {Goldhaber-Gordon}}]{hbn2}%
  \BibitemOpen
  \bibfield  {author} {\bibinfo {author} {\bibfnamefont {F.}~\bibnamefont
  {Amet}}, \bibinfo {author} {\bibfnamefont {J.~R.}\ \bibnamefont {Williams}},
  \bibinfo {author} {\bibfnamefont {K.}~\bibnamefont {Watanabe}}, \bibinfo
  {author} {\bibfnamefont {T.}~\bibnamefont {Taniguchi}}, \ and\ \bibinfo
  {author} {\bibfnamefont {D.}~\bibnamefont {Goldhaber-Gordon}},\ }\href@noop
  {} {\bibfield  {journal} {\bibinfo  {journal} {Phys. Rev. Lett.}\ }\textbf
  {\bibinfo {volume} {110}},\ \bibinfo {pages} {216601} (\bibinfo {year}
  {2013})}\BibitemShut {NoStop}%
\bibitem [{\citenamefont {Tang}\ \emph {et~al.}(2013)\citenamefont {Tang},
  \citenamefont {Wang}, \citenamefont {Zhang}, \citenamefont {Li},
  \citenamefont {Xie}, \citenamefont {Liu}, \citenamefont {Liu}, \citenamefont
  {Li}, \citenamefont {Huang}, \citenamefont {Xie},\ and\ \citenamefont
  {Jiang}}]{hbn3}%
  \BibitemOpen
  \bibfield  {author} {\bibinfo {author} {\bibfnamefont {S.}~\bibnamefont
  {Tang}}, \bibinfo {author} {\bibfnamefont {H.}~\bibnamefont {Wang}}, \bibinfo
  {author} {\bibfnamefont {Y.}~\bibnamefont {Zhang}}, \bibinfo {author}
  {\bibfnamefont {A.}~\bibnamefont {Li}}, \bibinfo {author} {\bibfnamefont
  {H.}~\bibnamefont {Xie}}, \bibinfo {author} {\bibfnamefont {X.}~\bibnamefont
  {Liu}}, \bibinfo {author} {\bibfnamefont {L.}~\bibnamefont {Liu}}, \bibinfo
  {author} {\bibfnamefont {T.}~\bibnamefont {Li}}, \bibinfo {author}
  {\bibfnamefont {F.}~\bibnamefont {Huang}}, \bibinfo {author} {\bibfnamefont
  {X.}~\bibnamefont {Xie}}, \ and\ \bibinfo {author} {\bibfnamefont
  {M.}~\bibnamefont {Jiang}},\ }\href@noop {} {\bibfield  {journal} {\bibinfo
  {journal} {Scientific Reports}\ }\textbf {\bibinfo {volume} {3}},\ \bibinfo
  {pages} {2666} (\bibinfo {year} {2013})}\BibitemShut {NoStop}%
\bibitem [{\citenamefont {Yang}\ \emph {et~al.}(2013)\citenamefont {Yang},
  \citenamefont {Chen}, \citenamefont {Shi}, \citenamefont {Liu}, \citenamefont
  {Zhang}, \citenamefont {Xie}, \citenamefont {Cheng}, \citenamefont {Wang},
  \citenamefont {Yang}, \citenamefont {Shi}, \citenamefont {Watanabe},
  \citenamefont {Taniguchi}, \citenamefont {Yao}, \citenamefont {Zhang},\ and\
  \citenamefont {Zhang}}]{hbn4}%
  \BibitemOpen
  \bibfield  {author} {\bibinfo {author} {\bibfnamefont {W.}~\bibnamefont
  {Yang}}, \bibinfo {author} {\bibfnamefont {G.}~\bibnamefont {Chen}}, \bibinfo
  {author} {\bibfnamefont {Z.}~\bibnamefont {Shi}}, \bibinfo {author}
  {\bibfnamefont {C.-C.}\ \bibnamefont {Liu}}, \bibinfo {author} {\bibfnamefont
  {L.}~\bibnamefont {Zhang}}, \bibinfo {author} {\bibfnamefont
  {G.}~\bibnamefont {Xie}}, \bibinfo {author} {\bibfnamefont {M.}~\bibnamefont
  {Cheng}}, \bibinfo {author} {\bibfnamefont {D.}~\bibnamefont {Wang}},
  \bibinfo {author} {\bibfnamefont {R.}~\bibnamefont {Yang}}, \bibinfo {author}
  {\bibfnamefont {D.}~\bibnamefont {Shi}}, \bibinfo {author} {\bibfnamefont
  {K.}~\bibnamefont {Watanabe}}, \bibinfo {author} {\bibfnamefont
  {T.}~\bibnamefont {Taniguchi}}, \bibinfo {author} {\bibfnamefont
  {Y.}~\bibnamefont {Yao}}, \bibinfo {author} {\bibfnamefont {Y.}~\bibnamefont
  {Zhang}}, \ and\ \bibinfo {author} {\bibfnamefont {G.}~\bibnamefont
  {Zhang}},\ }\href@noop {} {\bibfield  {journal} {\bibinfo  {journal} {Nature
  Mat.}\ }\textbf {\bibinfo {volume} {12}},\ \bibinfo {pages} {792} (\bibinfo
  {year} {2013})}\BibitemShut {NoStop}%
\bibitem [{\citenamefont {Tayran}\ \emph {et~al.}(2013)\citenamefont {Tayran},
  \citenamefont {Zhu}, \citenamefont {Baldoni}, \citenamefont {Selli},
  \citenamefont {Seifert},\ and\ \citenamefont {Tom{\'a}nek}}]{PRL2013}%
  \BibitemOpen
  \bibfield  {author} {\bibinfo {author} {\bibfnamefont {C.}~\bibnamefont
  {Tayran}}, \bibinfo {author} {\bibfnamefont {Z.}~\bibnamefont {Zhu}},
  \bibinfo {author} {\bibfnamefont {M.}~\bibnamefont {Baldoni}}, \bibinfo
  {author} {\bibfnamefont {D.}~\bibnamefont {Selli}}, \bibinfo {author}
  {\bibfnamefont {G.}~\bibnamefont {Seifert}}, \ and\ \bibinfo {author}
  {\bibfnamefont {D.}~\bibnamefont {Tom{\'a}nek}},\ }\href@noop {} {\bibfield
  {journal} {\bibinfo  {journal} {Phys. Rev. Lett.}\ }\textbf {\bibinfo
  {volume} {110}},\ \bibinfo {pages} {176805} (\bibinfo {year}
  {2013})}\BibitemShut {NoStop}%
\bibitem [{\citenamefont {Albrecht}\ and\ \citenamefont
  {Lyding}(2003)}]{Albrecht1}%
  \BibitemOpen
  \bibfield  {author} {\bibinfo {author} {\bibfnamefont {P.~M.}\ \bibnamefont
  {Albrecht}}\ and\ \bibinfo {author} {\bibfnamefont {J.~W.}\ \bibnamefont
  {Lyding}},\ }\href@noop {} {\bibfield  {journal} {\bibinfo  {journal} {Appl.
  Phys. Lett.}\ }\textbf {\bibinfo {volume} {83}},\ \bibinfo {pages} {5029}
  (\bibinfo {year} {2003})}\BibitemShut {NoStop}%
\bibitem [{\citenamefont {Orellana}, \citenamefont {Miwa},\ and\ \citenamefont
  {Fazzio}(2003)}]{Orellana}%
  \BibitemOpen
  \bibfield  {author} {\bibinfo {author} {\bibfnamefont {W.}~\bibnamefont
  {Orellana}}, \bibinfo {author} {\bibfnamefont {R.~H.}\ \bibnamefont {Miwa}},
  \ and\ \bibinfo {author} {\bibfnamefont {A.}~\bibnamefont {Fazzio}},\
  }\href@noop {} {\bibfield  {journal} {\bibinfo  {journal} {Phys. Rev. Lett.}\
  }\textbf {\bibinfo {volume} {91}},\ \bibinfo {pages} {166802} (\bibinfo
  {year} {2003})}\BibitemShut {NoStop}%
\bibitem [{\citenamefont {Barraza-Lopez}\ \emph {et~al.}(2006)\citenamefont
  {Barraza-Lopez}, \citenamefont {Albrecht}, \citenamefont {Romero},\ and\
  \citenamefont {Hess}}]{E1}%
  \BibitemOpen
  \bibfield  {author} {\bibinfo {author} {\bibfnamefont {S.}~\bibnamefont
  {Barraza-Lopez}}, \bibinfo {author} {\bibfnamefont {P.~M.}\ \bibnamefont
  {Albrecht}}, \bibinfo {author} {\bibfnamefont {N.~A.}\ \bibnamefont
  {Romero}}, \ and\ \bibinfo {author} {\bibfnamefont {K.}~\bibnamefont
  {Hess}},\ }\href@noop {} {\bibfield  {journal} {\bibinfo  {journal} {J. Appl.
  Phys.}\ }\textbf {\bibinfo {volume} {100}},\ \bibinfo {pages} {124304}
  (\bibinfo {year} {2006})}\BibitemShut {NoStop}%
\bibitem [{\citenamefont {Albrecht}, \citenamefont {Barraza-Lopez},\ and\
  \citenamefont {Lyding}(2007{\natexlab{a}})}]{E2}%
  \BibitemOpen
  \bibfield  {author} {\bibinfo {author} {\bibfnamefont {P.~M.}\ \bibnamefont
  {Albrecht}}, \bibinfo {author} {\bibfnamefont {S.}~\bibnamefont
  {Barraza-Lopez}}, \ and\ \bibinfo {author} {\bibfnamefont {J.~W.}\
  \bibnamefont {Lyding}},\ }\href@noop {} {\bibfield  {journal} {\bibinfo
  {journal} {Nanotechnology}\ }\textbf {\bibinfo {volume} {18}},\ \bibinfo
  {pages} {095204} (\bibinfo {year} {2007}{\natexlab{a}})}\BibitemShut
  {NoStop}%
\bibitem [{\citenamefont {Albrecht}, \citenamefont {Barraza-Lopez},\ and\
  \citenamefont {Lyding}(2007{\natexlab{b}})}]{E3}%
  \BibitemOpen
  \bibfield  {author} {\bibinfo {author} {\bibfnamefont {P.~M.}\ \bibnamefont
  {Albrecht}}, \bibinfo {author} {\bibfnamefont {S.}~\bibnamefont
  {Barraza-Lopez}}, \ and\ \bibinfo {author} {\bibfnamefont {J.~W.}\
  \bibnamefont {Lyding}},\ }\href@noop {} {\bibfield  {journal} {\bibinfo
  {journal} {Small}\ }\textbf {\bibinfo {volume} {3}},\ \bibinfo {pages} {1402}
  (\bibinfo {year} {2007}{\natexlab{b}})}\BibitemShut {NoStop}%
\bibitem [{\citenamefont {Barraza-Lopez}, \citenamefont {Albrecht},\ and\
  \citenamefont {Lyding}(2009)}]{E4}%
  \BibitemOpen
  \bibfield  {author} {\bibinfo {author} {\bibfnamefont {S.}~\bibnamefont
  {Barraza-Lopez}}, \bibinfo {author} {\bibfnamefont {P.~M.}\ \bibnamefont
  {Albrecht}}, \ and\ \bibinfo {author} {\bibfnamefont {J.~W.}\ \bibnamefont
  {Lyding}},\ }\href@noop {} {\bibfield  {journal} {\bibinfo  {journal} {Phys.
  Rev. B}\ }\textbf {\bibinfo {volume} {80}},\ \bibinfo {pages} {045415}
  (\bibinfo {year} {2009})}\BibitemShut {NoStop}%
\bibitem [{\citenamefont {Rivero}\ \emph {et~al.}(2015)\citenamefont {Rivero},
  \citenamefont {Horvath}, \citenamefont {Zhu}, \citenamefont {Guan},
  \citenamefont {Tom{\'a}nek},\ and\ \citenamefont
  {Barraza-Lopez}}]{USPRB2015}%
  \BibitemOpen
  \bibfield  {author} {\bibinfo {author} {\bibfnamefont {P.}~\bibnamefont
  {Rivero}}, \bibinfo {author} {\bibfnamefont {C.~M.}\ \bibnamefont {Horvath}},
  \bibinfo {author} {\bibfnamefont {Z.}~\bibnamefont {Zhu}}, \bibinfo {author}
  {\bibfnamefont {J.}~\bibnamefont {Guan}}, \bibinfo {author} {\bibfnamefont
  {D.}~\bibnamefont {Tom{\'a}nek}}, \ and\ \bibinfo {author} {\bibfnamefont
  {S.}~\bibnamefont {Barraza-Lopez}},\ }\href@noop {} {\bibfield  {journal}
  {\bibinfo  {journal} {Phys. Rev. B}\ }\textbf {\bibinfo {volume} {91}},\
  \bibinfo {pages} {115413} (\bibinfo {year} {2015})}\BibitemShut {NoStop}%
\bibitem [{\citenamefont {Soler}\ \emph {et~al.}(2002)\citenamefont {Soler},
  \citenamefont {Artacho}, \citenamefont {Gale}, \citenamefont {Garc{\'\i}a},
  \citenamefont {Junquera}, \citenamefont {Ordej{\'o}n},\ and\ \citenamefont
  {S{\'a}nchez-Portal}}]{SIESTA1}%
  \BibitemOpen
  \bibfield  {author} {\bibinfo {author} {\bibfnamefont {J.~M.}\ \bibnamefont
  {Soler}}, \bibinfo {author} {\bibfnamefont {E.}~\bibnamefont {Artacho}},
  \bibinfo {author} {\bibfnamefont {J.~D.}\ \bibnamefont {Gale}}, \bibinfo
  {author} {\bibfnamefont {A.}~\bibnamefont {Garc{\'\i}a}}, \bibinfo {author}
  {\bibfnamefont {J.}~\bibnamefont {Junquera}}, \bibinfo {author}
  {\bibfnamefont {P.}~\bibnamefont {Ordej{\'o}n}}, \ and\ \bibinfo {author}
  {\bibfnamefont {D.}~\bibnamefont {S{\'a}nchez-Portal}},\ }\href@noop {}
  {\bibfield  {journal} {\bibinfo  {journal} {J. Phys.: Condens. Matter}\
  }\textbf {\bibinfo {volume} {14}},\ \bibinfo {pages} {2745} (\bibinfo {year}
  {2002})}\BibitemShut {NoStop}%
\bibitem [{\citenamefont {Artacho}\ \emph {et~al.}(2008)\citenamefont
  {Artacho}, \citenamefont {Anglada}, \citenamefont {Di{\'e}guez},
  \citenamefont {Gale}, \citenamefont {Garc{\'\i}a}, \citenamefont {Junquera},
  \citenamefont {Martin}, \citenamefont {Ordej{\'o}n}, \citenamefont {Pruneda},
  \citenamefont {S{\'a}nchez-Portal},\ and\ \citenamefont {Soler}}]{SIESTA2}%
  \BibitemOpen
  \bibfield  {author} {\bibinfo {author} {\bibfnamefont {E.}~\bibnamefont
  {Artacho}}, \bibinfo {author} {\bibfnamefont {E.}~\bibnamefont {Anglada}},
  \bibinfo {author} {\bibfnamefont {O.}~\bibnamefont {Di{\'e}guez}}, \bibinfo
  {author} {\bibfnamefont {J.~D.}\ \bibnamefont {Gale}}, \bibinfo {author}
  {\bibfnamefont {A.}~\bibnamefont {Garc{\'\i}a}}, \bibinfo {author}
  {\bibfnamefont {J.}~\bibnamefont {Junquera}}, \bibinfo {author}
  {\bibfnamefont {R.~M.}\ \bibnamefont {Martin}}, \bibinfo {author}
  {\bibfnamefont {P.}~\bibnamefont {Ordej{\'o}n}}, \bibinfo {author}
  {\bibfnamefont {J.~M.}\ \bibnamefont {Pruneda}}, \bibinfo {author}
  {\bibfnamefont {D.}~\bibnamefont {S{\'a}nchez-Portal}}, \ and\ \bibinfo
  {author} {\bibfnamefont {J.~M.}\ \bibnamefont {Soler}},\ }\href@noop {}
  {\bibfield  {journal} {\bibinfo  {journal} {J. Phys.: Condens. Matter}\
  }\textbf {\bibinfo {volume} {20}},\ \bibinfo {pages} {064208} (\bibinfo
  {year} {2008})}\BibitemShut {NoStop}%
\bibitem [{\citenamefont {Ceperley}\ and\ \citenamefont {Alder}(1980)}]{LDA1}%
  \BibitemOpen
  \bibfield  {author} {\bibinfo {author} {\bibfnamefont {D.~M.}\ \bibnamefont
  {Ceperley}}\ and\ \bibinfo {author} {\bibfnamefont {B.~J.}\ \bibnamefont
  {Alder}},\ }\href@noop {} {\bibfield  {journal} {\bibinfo  {journal} {Phys.
  Rev. Lett.}\ }\textbf {\bibinfo {volume} {45}},\ \bibinfo {pages} {566}
  (\bibinfo {year} {1980})}\BibitemShut {NoStop}%
\bibitem [{\citenamefont {Perdew}\ and\ \citenamefont {Zunger}(1981)}]{LDA2}%
  \BibitemOpen
  \bibfield  {author} {\bibinfo {author} {\bibfnamefont {J.~P.}\ \bibnamefont
  {Perdew}}\ and\ \bibinfo {author} {\bibfnamefont {A.}~\bibnamefont
  {Zunger}},\ }\href@noop {} {\bibfield  {journal} {\bibinfo  {journal} {Phys.
  Rev. B}\ }\textbf {\bibinfo {volume} {23}},\ \bibinfo {pages} {5048}
  (\bibinfo {year} {1981})}\BibitemShut {NoStop}%
\bibitem [{\citenamefont {Wang}\ \emph {et~al.}(2004)\citenamefont {Wang},
  \citenamefont {Gao}, \citenamefont {Guo}, \citenamefont {Liu}, \citenamefont
  {Batyrev}, \citenamefont {McMahon},\ and\ \citenamefont {Zhang}}]{PRB2004}%
  \BibitemOpen
  \bibfield  {author} {\bibinfo {author} {\bibfnamefont {Y.~L.}\ \bibnamefont
  {Wang}}, \bibinfo {author} {\bibfnamefont {H.-J.}\ \bibnamefont {Gao}},
  \bibinfo {author} {\bibfnamefont {H.~M.}\ \bibnamefont {Guo}}, \bibinfo
  {author} {\bibfnamefont {H.~W.}\ \bibnamefont {Liu}}, \bibinfo {author}
  {\bibfnamefont {I.~G.}\ \bibnamefont {Batyrev}}, \bibinfo {author}
  {\bibfnamefont {W.~E.}\ \bibnamefont {McMahon}}, \ and\ \bibinfo {author}
  {\bibfnamefont {S.~B.}\ \bibnamefont {Zhang}},\ }\href@noop {} {\bibfield
  {journal} {\bibinfo  {journal} {Phys. Rev. B}\ }\textbf {\bibinfo {volume}
  {70}},\ \bibinfo {pages} {073312} (\bibinfo {year} {2004})}\BibitemShut
  {NoStop}%
\end{thebibliography}


\end{document}